\documentclass[preprint,10pt,1p,sort&compress]{elsarticle}
\pdfoutput=1
\usepackage{amssymb,amsmath}
\usepackage{latexsym,mathrsfs}
\usepackage{graphicx,subfigure}
\usepackage{epsfig,varioref}
\usepackage{xr-hyper,color,multirow}
\usepackage{tikz,array,wasysym}
\usepackage{color,float}
\usepackage[utf8]{inputenc}

\journal{Phys. Dark Univ.}

\begin{document}

\begin{frontmatter}

\title{Interacting dark energy in the early 2020s: a promising solution to the $H_0$ and cosmic shear tensions}

\author[manchester]{Eleonora Di Valentino}
\ead{eleonora.divalentino@manchester.ac.uk}

\author[rome]{Alessandro Melchiorri}
\ead{alessandro.melchiorri@roma1.infn.it}

\author[ific]{Olga Mena}
\ead{omena@ific.uv.es}

\author[kicc]{Sunny Vagnozzi}
\ead{sunny.vagnozzi@ast.cam.ac.uk}

\address[manchester]{Jodrell Bank Center for Astrophysics, University of Manchester, Manchester M13 9PL, UK}
\address[rome]{Department of Physics and INFN, University of Rome ``La Sapienza'', 00185 Rome, Italy}
\address[ific]{Instituto de F\'{i}sica Corpuscular, University of Valencia-CSIC, 46980 Valencia, Spain}
\address[kicc]{Kavli Institute for Cosmology, University of Cambridge, Cambridge CB3 0HA, UK}

\begin{abstract}
\noindent We examine interactions between dark matter and dark energy in light of the latest cosmological observations, focusing on a specific model with coupling proportional to the dark energy density. Our data includes Cosmic Microwave Background (CMB) measurements from the \textit{Planck} 2018 legacy data release, late-time measurements of the expansion history from Baryon Acoustic Oscillations (BAO) and Supernovae Type Ia (SNeIa), galaxy clustering and cosmic shear measurements from the \textit{Dark Energy Survey} Year 1 results, and the 2019 local distance ladder measurement of the Hubble constant $H_0$ from the \textit{Hubble Space Telescope}. Considering \textit{Planck} data both in combination with BAO or SNeIa data reduces the $H_0$ tension to a level which could possibly be compatible with a statistical fluctuation. The very same model also significantly reduces the $\Omega_{\rm m}-\sigma_8$ tension between CMB and cosmic shear measurements. Interactions between the dark sectors of our Universe remain therefore a promising joint solution to these persisting cosmological tensions.
\end{abstract}


\end{frontmatter}

\section{Introduction}
\label{sec:intro}

The concordance $\Lambda$CDM cosmological model has been incredibly successful at describing cosmological observations at high and low redshift~\cite{Riess:1998cb,Perlmutter:1998np,Ade:2015xua,Alam:2016hwk,Troxel:2017xyo}. Yet, as uncertainties on cosmological parameters keep shrinking, a number of weaknesses have emerged: one of the most intriguing ones is the ``$H_0$ tension'', referring to the mismatch between the value of the Hubble constant $H_0$ inferred from \textit{Planck} Cosmic Microwave Background (CMB) data and direct local distance ladder measurements~\cite{Freedman:2017yms,DiValentino:2017gzb}. In the past decade we have witnessed the tension between these two values grow in significance level from $2\sigma$ to $4.4\sigma$: the latest determinations from the \textit{Planck} 2018 results and from the observations of Large Magellanic Cloud Cepheids by the \textit{Hubble Space Telescope} (\textit{HST}; measurement denoted as R19 hereafter) give $h = (0.6737 \pm 0.0054)$~\cite{Aghanim:2018eyx} and $h=(0.7403 \pm 0.0142)$~\cite{Riess:2019cxk} respectively, with $h \equiv H_0/(100\,{\rm km}\,{\rm s}^{-1}\,{\rm Mpc}^{-1})$ the reduced Hubble constant. A very appealing possibility is that the $H_0$ discrepancy might be a hint of physics beyond the canonical $\Lambda$CDM model. The most economic possibilities in this direction involve phantom dark energy or some form of dark radiation, but a number of more complex scenarios have been studied, e.g.~\cite{Alam:2016wpf,DiValentino:2016hlg,Qing-Guo:2016ykt,Bernal:2016gxb,Ko:2016uft,Karwal:2016vyq,Chacko:2016kgg,Zhao:2017cud,Vagnozzi:2017ovm,Agrawal:2017rvu,Benetti:2017gvm,Feng:2017nss,Zhao:2017urm,DiValentino:2017zyq,Gariazzo:2017pzb,Dirian:2017pwp,Sola:2017znb,Feng:2017mfs,Renk:2017rzu,Yang:2017alx,Buen-Abad:2017gxg,Raveri:2017jto,DiValentino:2017rcr,DiValentino:2017oaw,Khosravi:2017hfi,Peirone:2017vcq,Benetti:2017juy,Mortsell:2018mfj,Gomez-Valent:2018nib,Vagnozzi:2018jhn,Nunes:2018xbm,Poulin:2018zxs,Kumar:2018yhh,Banihashemi:2018oxo,DEramo:2018vss,Guo:2018ans,Graef:2018fzu,Yang:2018qmz,Banihashemi:2018has,Aylor:2018drw,Poulin:2018cxd,Kreisch:2019yzn,Pandey:2019plg,Vattis:2019efj,Colgain:2019pck,Agrawal:2019lmo,Li:2019san,Yang:2019jwn,Rezaei:2019xwo,Keeley:2019esp,Li:2019ypi,Rossi:2019lgt,DiValentino:2019exe,Archidiacono:2019wdp,Desmond:2019ygn,Yang:2019nhz,Nesseris:2019fwr,Vagnozzi:2019ezj,Visinelli:2019qqu,Cai:2019bdh,Schoneberg:2019wmt,Pan:2019hac,DiValentino:2019dzu,Xiao:2019ccl,Panpanich:2019fxq,Knox:2019rjx,Sola:2019jek,Escudero:2019gvw,Yan:2019gbw,Banerjee:2019kgu,Cheng:2019bkh,Liu:2019awo,Anchordoqui:2019amx,Frusciante:2019puu,Akarsu:2019hmw,Ding:2019mmw,Ye:2020btb,Yang:2020zuk,Perez:2020cwa,Pan:2020mst,Choi:2020tqp,Pan:2020bur,Krishnan:2020obg,DAgostino:2020dhv,Nunes:2020hzy,Hogg:2020rdp,Heinesen:2020sre,Benevento:2020fev,Zumalacarregui:2020cjh,Desmond:2020wep,Akarsu:2020yqa,Haridasu:2020xaa,Alestas:2020mvb,Jedamzik:2020krr,Braglia:2020iik,Chudaykin:2020acu,Jimenez:2020bgw,Ballardini:2020iws,Anchordoqui:2020znj,Banerjee:2020xcn}.

On  the  other  hand, tensions between cosmic shear surveys (such as~\cite{Kohlinger:2017sxk,Hildebrandt:2016iqg,Joudaki:2016kym}) and CMB measurements have also emerged~\cite{Hildebrandt:2016iqg,Joudaki:2017zdt,DiValentino:2018gcu}. For instance, the quantity $S_8 \equiv \sigma_8 (\Omega_{\rm m}/0.3)^{0.5}$ as measured by the KiDS weak lensing survey was shown to be in $2.6\sigma$ tension with the same quantity as measured by \textit{Planck}~\cite{Joudaki:2017zdt,Hildebrandt:2016iqg} (see also~\cite{Kilbinger:2012qz,Heymans:2013fya} for previous analyses of CFHTLenS data). Focusing on the joint galaxy clustering and lensing likelihoods from the Dark Energy Survey (\textit{DES})~\cite{Troxel:2017xyo,Abbott:2017wau,Krause:2017ekm}, the \textit{Planck} collaboration found modest tension with the \textit{DES} results when galaxy clustering measurements are included, as the latter prefer an $\approx 2.5\sigma$ lower value of $S_8$~\cite{Aghanim:2018eyx}.

It is worth clarifying that the situation concerning the $S_8$/cosmic shear tension is less well defined with respect to that of the $H_0$ tension. In fact, various weak lensing analyses report varying degrees of tension. Earlier we have discussed the analyses which show strongest tension with \textit{Planck}. However, other analyses show no tension, or only a mild amount of tension: these include for instance a combination of KiDS and GAMA galaxy clustering~\cite{vanUitert:2017ieu}, a re-analysis of KiDS after accounting for a more careful treatment of intrinsic galaxy shape noise, multiplicative shear calibration uncertainties, and bin angular scale~\cite{Troxel:2018qll}, and \textit{DES} results when only considering cosmic shear but not galaxy clustering and shear-galaxy cross-correlations~\cite{Troxel:2017xyo,Abbott:2017wau,Krause:2017ekm}. Despite these uncertainties, a number of exotic scenarios have been advocated in the past to alleviate the $S_8$ tension, see for instance~\cite{Joudaki:2016kym,Ko:2016uft,Chacko:2016kgg,Pourtsidou:2016ico,Buen-Abad:2017gxg,DiValentino:2017oaw,Benetti:2017juy,Sola:2017jbl,Gomez-Valent:2017idt,Gariazzo:2017pzb,Barros:2018efl,Poulin:2018zxs,Kreisch:2019yzn,Keeley:2019esp,Kazantzidis:2018rnb,Kazantzidis:2019dvk,Benisty:2020kdt}.

Within the $\Lambda$CDM model, dark matter (DM) and dark energy (DE) behave as separate fluids not sharing interactions beyond gravitational ones. Historically, interactions between DM and DE were originally introduced to alleviate the coincidence problem~\cite{Amendola:1999er,Mangano:2002gg}, although it is now understood that the energy exchange required for this problem to be addressed by these means is too large and observationally excluded. Nonetheless, interactions between DM and DE cannot be excluded on general grounds, and it is thus worth constraining them against the available wealth of precision cosmological data. This has motivated a large number of studies based on models where DM and DE share interactions, usually referred to as interacting dark energy (IDE) models (see e.g.~\cite{Wetterich:1994bg,Amendola:1999dr,Amendola:2003eq,Farrar:2003uw,Amendola:2004ew,Pettorino:2004zt,Pettorino:2005pv,Barrow:2006hia,Mainini:2006zj,Amendola:2006dg,Mainini:2007ft,Bean:2007ny,He:2008tn,Pettorino:2008ez,Valiviita:2008iv,Vergani:2008jv,Baldi:2008ay,Gavela:2009cy,LaVacca:2009yp,CalderaCabral:2009ja,Majerotto:2009np,Abdalla:2009mt,Saracco:2009df,Wintergerst:2010ui,Honorez:2010rr,Baldi:2010td,Baldi:2010ks,Baldi:2010pq,Baldi:2011wa,Baldi:2011th,Baldi:2011qi,Clemson:2011an,Lee:2011tq,Marulli:2011jk,Amendola:2011ie,Baldi:2011wy,Beynon:2011hw,Cui:2012is,Bonometto:2012qz,Pettorino:2012ts,Pan:2012ki,Giocoli:2013ba,Salvatelli:2013wra,Carbone:2013dna,Piloyan:2013mla,Pettorino:2013oxa,Pourtsidou:2013nha,Pace:2013pea,Bonometto:2013eva,Yang:2014gza,Piloyan:2014gta,Yang:2014vza,Baldi:2014tja,Nunes:2014qoa,Faraoni:2014vra,Amendola:2014kwa,Pan:2014afa,Ferreira:2014cla,Ade:2015rim,Skordis:2015yra,Maccio:2015iya,Bonometto:2015mya,Penzo:2015tha,Tamanini:2015iia,Fontanot:2015xoa,Li:2015vla,Pollina:2015uaa,Casas:2015qpa,Odderskov:2015fba,Murgia:2016ccp,Wang:2016lxa,Nunes:2016dlj,Yang:2016evp,Pan:2016ngu,Sharov:2017iue,Benisty:2017eqh,Bonometto:2017rdu,An:2017kqu,Miranda:2017rdk,Santos:2017bqm,Mifsud:2017fsy,Kumar:2017bpv,Guo:2017deu,Pan:2017ent,Linton:2017ged,An:2017crg,Dutta:2017fjw,Benisty:2018qed,Costa:2018aoy,Wang:2018azy,Hashim:2018dek,vonMarttens:2018iav,Bonometto:2018dmx,Benisty:2018oyy,Yang:2018qec,Costa:2019uvk,Martinelli:2019dau,Li:2019loh,Yang:2019vni,Bachega:2019fki,Li:2019ajo,Mukhopadhyay:2019jla,Carneiro:2019rly,Kase:2019veo,Liu:2019ygl,Kase:2019mox,Chamings:2019kcl,Benisty:2020nuu,Amendola:2020ldb,Benisty:2020nql}, for a recent comprehensive review see~\cite{Wang:2016lxa}). Several studies in the literature have been devoted to exploring whether DM-DE interactions may help resolve the enduring $H_0$ tension, see e.g.~\cite{Salvatelli:2014zta,Kumar:2016zpg,Xia:2016vnp,Kumar:2017dnp,DiValentino:2017iww,Yang:2017ccc,Feng:2017usu,Yang:2018ubt,Yang:2018xlt,Yang:2018uae,Li:2018ydj,Kumar:2019wfs,Pan:2019jqh,Yang:2019uzo,Pan:2019gop,Benetti:2019lxu,Yang:2019uog,DiValentino:2019jae,Gomez-Valent:2020mqn,Aljaf:2020eqh}.

In this work we (re)assess whether IDE cosmologies still provide a viable solution to the $H_0$ tension in light of the latest \textit{Planck} and \textit{HST} measurements. We find that IDE provides an interesting solution to the $H_0$ tension, which is brought below the $1\sigma$ level when considering \textit{Planck} data alone. Intriguingly, when combining the latest \textit{Planck} and \textit{HST} measurements we find very strong indications for an interaction between the two dark components. These findings are, however, softened when including late-time measurements of the expansion history from Baryon Acoustic Oscillations and Supernovae Type Ia. We find that IDE also provides a promising solution to the $S_8$ tension between \textit{Planck} and \textit{DES}.

The rest of this paper is organized as follows. In Sec.~\ref{sec:ide} we revisit the basic background and perturbation equations of the IDE model we consider, discussing also stability considerations. In Sec.~\ref{sec:methodology} we discuss the analysis method adopted, as well as the cosmological datasets considered. Our results are presented in Sec.~\ref{sec:results}, where we argue that the IDE model considered can partially address the $H_0$ and cosmic shear tensions, whereas possible model-dependence issues pertaining to the use of BAO and SNeIa data when constraining IDE models and assessing their ability to address the $H_0$ tension are discussed in Sec.~\ref{subsec:baosne}. Finally, we provide closing remarks in Sec.~\ref{sec:conclusions}.

\section{Interacting dark energy revisited}
\label{sec:ide}

We consider model featuring interactions between DM and DE with energy exchange proportional to the DM four-velocity, studied in earlier works such as~\cite{Valiviita:2008iv,delCampo:2008jx,Gavela:2009cy,Honorez:2010rr}. We assume a pressureless cold DM component and a DE component with equation of state (EoS) $w$, and denote the DM and DE energy densities by $\rho_c$ and $\rho_x$ respectively. At the background level, the DM-DE coupling modifies the continuity equations for the two dark fluids as follows~\cite{Gavela:2009cy}:
\begin{eqnarray}
\label{eq:continuitydensityc}
\dot{\rho}_c+3{\cal H}\rho_c &=& Q\,, \\
\label{eq:continuitydensityx}
\dot{\rho}_x+3{\cal H}(1+w)\rho_x &=&-Q\,,
\end{eqnarray}
where the dot denotes derivative with respect to conformal time $\tau$, and ${\cal H} \equiv \dot{a}/a$ is the conformal Hubble rate. In the notation of Eqs.~(\ref{eq:continuitydensityc},\ref{eq:continuitydensityx}), $Q>0$ and $Q<0$ indicate energy transfer from DE to DM and viceversa. We choose to focus on an interacting dark energy model, studied in various earlier works, wherein the coupling $Q$ takes the following form~\cite{Valiviita:2008iv,Gavela:2009cy}:
\begin{eqnarray}
Q = \xi{\cal H}\rho_x\,,
\label{eq:coupling}
\end{eqnarray}
where $\xi$ is a dimensionless coupling governing the strength of the DM-DE interaction. We shall refer to the resulting model as $\xi\Lambda$CDM model, or coupled vacuum model.

The presence of the DM-DE coupling also modifies the evolution of perturbations. In synchronous gauge, the linear perturbation equations for the evolution of the DM and DE density perturbations $\delta$ and velocity divergences $\theta$ are given by~\cite{Valiviita:2008iv,Gavela:2009cy,Gavela:2010tm}:
\begin{eqnarray}
\label{eq:deltac}
\dot{\delta}_c &=& -\theta_c - \frac{1}{2}\dot{h} +\xi{\cal H}\frac{\rho_x}{\rho_c}(\delta_x-\delta_c)+\xi\frac{\rho_x}{\rho_c} \left ( \frac{kv_T}
{3}+\frac{\dot{h}}{6} \right )\,, \\
\label{eq:thetac}
\dot{\theta}_c &=& -{\cal H}\theta_c\,,\\
\label{eq:deltax}
\dot{\delta}_x &=& -(1+w) \left ( \theta_x+\frac{\dot{h}}{2} \right )-\xi \left ( \frac{kv_T}{3}+\frac{\dot{h}}{6} \right ) \nonumber \\
&&-3{\cal H}(1-w) \left [ \delta_x+\frac{{\cal H}\theta_x}{k^2} \left (3(1+w)+\xi \right ) \right ]\,,\\
\label{eq:thetax}
\dot{\theta}_x &=& 2{\cal H}\theta_x+\frac{k^2}{1+w}\delta_x+2{\cal H}\frac{\xi}{1+w}\theta_x-\xi{\cal H}\frac{\theta_c}{1+w}\,.
\end{eqnarray}
It is worth pointing out that the specific coupling chosen in Eq.~(\ref{eq:coupling}) is purely phenomenological, introduced at the level of the continuity equations as shown in Eq.~(\ref{eq:continuitydensityc},\ref{eq:continuitydensityx}). In other words, it is not derived from a specific action (\textit{e.g.} as in specific coupled quintessence models), and its phenomenological nature should be kept in mind.~\footnote{Nonetheless, given a certain field model of DM and DE, it is in principle possible to work backwards and reconstruct the specific type of Lagrangian interaction term between the DM and DE fields which could give rise to a chosen coupling function $Q$, as was done recently for instance in~\cite{Pan:2020zza}. However, there is no guarantee that this procedure would return a Lagrangian term which is well-motivated from first-principle field-theoretical considerations.} This explains for instance the difference between our results and those of the \textit{Planck} collaboration, who in their 2015 dark energy and modified gravity paper~\cite{Ade:2015rim} also considered constraints on coupled DE models. In~\cite{Ade:2015rim} the Planck collaboration studied coupled quintessence models featuring an exponential coupling between a DM field and a scalar DE field (such a coupling could be motivated by Weyl scaling scalar-tensor theories), with an inverse power-law potential for the latter. The model and corresponding coupling functions were based on a specific well-motivated Lagrangian, whereas our choice of coupling function is phenomenological.

When introducing DM-DE interactions, we also need to modify the initial conditions for our Boltzmann system appropriately. To do so, we follow the earlier work of~\cite{Gavela:2010tm} and set adiabatic initial conditions for all species, which in turn requires appropriately modifying the initial conditions for $\delta_x$ and $\theta_x$. As shown in~\cite{Gavela:2010tm}, following the gauge invariant formulation of~\cite{Doran:2003xq}, one can identify the appropriate initial conditions for the density contrast and velocity divergence of each species by writing the coupled Boltzmann equations in matrix form, solving the appropriate eigenvalue/eigenvector problem, and studying the modes which dominate the time evolution (which in this case will be those corresponding to the largest eigenvalue). These initial conditions turn out to be (see e.g.~\cite{Gavela:2010tm,Salvatelli:2013wra,DiValentino:2017iww,DiValentino:2019jae,Lucca:2020zjb}):
\begin{eqnarray}
\delta_x^{\rm in}(\eta) &=& \frac{3}{2}\frac{(2\xi-1-w)(1+w+\xi/3)}{12w^2-2w-3w\xi+7\xi-14}\delta_{\gamma}^{\rm in}(\eta)\,,\\
\theta_x^{\rm in}(x) &=& \frac{3}{2}\frac{k\eta(1+w+\xi/3)}{2w+3w\xi+14-12w^2-7\xi}\delta_{\gamma}^{\rm in}(\eta)\,,
\label{eq:initialconditions}
\end{eqnarray}
where $\eta= k \tau$ and $\delta_{\gamma}^{\rm in}(\eta)$ denotes the initial conditions for the photon density perturbations. We therefore set the initial conditions for the DE density contrast and velocity divergence following Eq.~(\ref{eq:initialconditions}).

In the presence of DM-DE interactions, care must be given to the stability of the interacting system. For $w=-1$ (\textit{i.e.} interacting vacuum), IDE models can suffer from gravitational instabilities~\cite{Valiviita:2008iv,He:2008si}. However, even when $w \neq -1$, one has to worry about early-time instabilities, leading to curvature perturbations blowing up on superhorizon scales. For the $\xi\Lambda$CDM model in which $Q \propto \rho_x$, the instabilities present when $w \neq -1$ are absent if the signs of $\xi$ and $(1+w)$ are opposite~\cite{Valiviita:2008iv,He:2008si,Jackson:2009mz, Gavela:2010tm,Clemson:2011an} (see also~\cite{Li:2014eha,Li:2014cee,Guo:2017hea,Zhang:2017ize,Guo:2018gyo,Yang:2018euj,Dai:2019vif} for alternative approaches to avoiding these instabilities).

\section{Methodology and cosmological observations}
\label{sec:methodology}

We consider an IDE model characterized by the coupling given by Eq.~(\ref{eq:coupling}). The model is described by the usual six cosmological parameters of $\Lambda$CDM ($\Omega_{\rm b}h^2$, $\Omega_{\rm c}h^2$, $\theta_{\rm s}$, $A_{\rm s}$, $n_{\rm s}$, and $\tau$), in addition to the DM-DE coupling $\xi$. To circumvent the instability problem discussed above, we fix the DE EoS to $w=-0.999$. The rationale behind this approach (already followed in~\cite{Salvatelli:2013wra,DiValentino:2017iww}) is that for $w$ sufficiently close to $-1$ the effect of DE perturbations in Eqs.~(\ref{eq:deltax},\ref{eq:thetax}) is basically unnoticeable: consequently, these equations are essentially only capturing the effect of the DM-DE coupling $\xi$, while at the same time ensuring the absence of gravitational instabilities present when $w$ is strictly equal to $-1$. Such a model provides therefore a rather accurate surrogate for a  $\Lambda$CDM+$\xi$ cosmology, and we shall refer to this model as $\xi\Lambda$CDM model (or coupled vacuum model). In order to avoid early-time instabilities discussed above, we then need to impose that $(1+w)$ and $\xi$ have opposite signs. Since we fixed $w=-0.999$, we impose $\xi<0$, and are therefore considering a model where energy flows from DM to DE. In a more realistic analysis, one should also vary the DE EoS (paying attention to the instability issue, which requires $\xi$ to change sign depending on whether the DE EoS lies in the quintessence or phantom regime), as we have done in our later companion paper~\cite{DiValentino:2019jae}.

Data-wise, we first consider measurements of CMB temperature and polarization anisotropies, as well as their cross-correlations, from the \textit{Planck} 2018 legacy data release~\cite{Aghanim:2018eyx,Aghanim:2018oex}. This dataset is referred to as Planck TT,TE,EE+lowE in~\cite{Aghanim:2018eyx}, whereas we refer to it simply as \textit{Planck}. We then include measurements of the CMB lensing power spectrum reconstructed from the CMB temperature four-point function~\cite{Aghanim:2018oex}. We refer to this dataset as \textit{lensing}.

In addition to CMB data, we then consider Baryon Acoustic Oscillation (BAO) measurements from the 6dFGS~\cite{Beutler:2011hx}, SDSS-MGS~\cite{Ross:2014qpa}, and BOSS DR12~\cite{Alam:2016hwk} surveys, a combination which we refer to as \textit{BAO}. We further include Type Ia Supernovae (SNeIa) distance moduli measurements from the \textit{Pantheon} sample~\cite{Scolnic:2017caz}, referring to this dataset as \textit{Pantheon}. As pointed out in a number of earlier works~\cite{Bernal:2016gxb,Feeney:2018mkj,Lemos:2018smw,Aylor:2018drw,Knox:2019rjx}, it is important to consider BAO and SNeIa distance measurements, from which one can construct an inverse distance ladder, when constraining late-time deviations from $\Lambda$CDM. In fact, studies based on the inverse distance ladder approach suggest that finding late-time solutions to the $H_0$ tension which can fit BAO and SNe data is challenging (albeit not impossible), see e.g. Fig.~1 in~\cite{Knox:2019rjx}.

We also include galaxy clustering and cosmic shear measurements, as well as their cross-correlations, from the Dark Energy Survey combined-probe Year 1 results~\cite{Troxel:2017xyo,Abbott:2017wau,Krause:2017ekm}. In particular, we consider the shear-shear, galaxy-galaxy, and galaxy-shear correlation functions, with the combination often referred to as the ``$3 \times 2$pt likelihood''. We refer to this dataset as \textit{DES}.

Finally, we also consider a Gaussian prior on the Hubble constant $H_0=74.03 \pm 1.42\,{\rm km}\,{\rm s}^{-1}\,{\rm Mpc}^{-1}$, consistent with the latest measurement by \textit{HST} in~\cite{Riess:2019cxk}. We refer to this prior as \textit{R19}. We focus on the ability of our model to reduce the tension with the \textit{R19} measurement (and subsequently consider the corresponding prior) as this is the measurement which is most discrepant with the \textit{Planck}+\textit{BAO} dataset combination assuming $\Lambda$CDM, and hence sets a harder task for our model. However, we note that there are alternative local measurements of $H_0$ which are also discrepant with the same dataset combination with various degrees of tension, including: local SNeIa calibrated with the Tip of the Red Giant Branch~\cite{Freedman:2019jwv}, strong lensing time delays from distant quasars as measured e.g. by the H0LiCOW team~\cite{Wong:2019kwg}, and water megamasers~\cite{Pesce:2020xfe}. See e.g. Fig.~1 in~\cite{Verde:2019ivm} for a visual summary of some of the main local measurements of $H_0$ beyond \textit{R19}.

We modify the Boltzmann solver \texttt{CAMB}~\cite{Lewis:1999bs} to incorporate the effect of the DM-DE coupling as in Eqs.~(\ref{eq:deltac},\ref{eq:thetax}). We sample the posterior distribution of the cosmological parameters by making use of Markov Chain Monte Carlo (MCMC) methods, through a modified version of the publicly available MCMC sampler \texttt{CosmoMC}~\cite{Lewis:2002ah}. We monitor the convergence of the generated MCMC chains through the Gelman-Rubin parameter $R-1$~\cite{Gelman:1992zz}, requiring $R-1<0.02$ for our MCMC chains to be considered as converged. We impose flat priors on all cosmological parameters unless otherwise stated. In particular, as required by stability considerations, we impose $\xi<0$ at the prior level.

Finally, we use our MCMC chains to compute the Bayesian evidence for the $\xi\Lambda$CDM model (for different choices of datasets) using the \texttt{MCEvidence} code~\cite{Heavens:2017afc}. We then compute the (logarithm of the) Bayes factor with respect to $\Lambda$CDM, $\ln B$, with a value $\ln B>0$ indicating that the $\xi\Lambda$CDM model is preferred. We qualify the strength of the obtained values of $\ln B$ using the modified version of the Jeffreys scale provided in~\cite{Kass:1995loi}.
\begin{table}[!h]
\centering                                         
\begin{tabular}{ccccc}       
\hline\hline                                                                                                                    
Parameter & \textit{Planck} & \textit{Planck}+\textit{lensing} & \textit{Planck}+\textit{BAO} & \textit{Planck}+\textit{Pantheon} \\ \hline

$\Omega_{\rm b}h^2$ & $0.0224 \pm 0.0002$ & $0.0224 \pm 0.0002$ & $0.0224 \pm 0.0002$ & $0.0224 \pm 0.0002$ \\
 
$\Omega_{\rm c}h^2$ & $<0.105$ & $<0.108$ & $0.095^{+0.022}_{-0.008}$ & $0.103^{+0.013}_{-0.007}$ \\

$n_s$ & $0.966 \pm 0.004$ & $0.966 \pm 0.004$ & $0.965 \pm 0.004$ & $0.964 \pm 0.004$ \\

$\tau$ & $0.054 \pm 0.008$ & $0.053 \pm 0.007$ & $0.054 \pm 0.008$ & $0.054 \pm 0.008$ \\

$\xi$ & $-0.54^{+0.12}_{-0.28}$ & $-0.51^{+0.12}_{-0.29}$ & $-0.22^{+0.21}_{-0.05}$ & $-0.15^{+0.12}_{-0.07}$ \\

$H_0\,[{\rm km}\,{\rm s}^{-1}\,{\rm Mpc}^{-1}]$ & $72.8^{+3.0}_{-1.5}$ & $72.8^{+3.3}_{-1.6}$ & $69.4^{+0.9}_{-1.5}$ & $68.6^{+0.8}_{-1.0}$ \\

\hline\hline                                                  
\end{tabular}                                                   
\caption{Constraints on selected cosmological parameters within the coupled vacuum $\xi\Lambda$CDM model, considering either the \textit{Planck} 2018 legacy dataset alone, or the same dataset in combination with: the CMB lensing power spectrum reconstructed from the CMB temperature four-point function; a combination of Baryon Acoustic Oscillation distance measurements; and distance moduli measurements from the \textit{Pantheon} Supernovae Type Ia catalogue. Constraints are reported as 68\%~C.L. intervals. In the case of $\Omega_{\rm c}h^2$ for the \textit{Planck} and \textit{Planck}+\textit{lensing} dataset combinations the quantity quoted is the 95\%~C.L. upper limit.}
\label{tab7p} 
\end{table}   

\begin{table}[!h]
\centering                                         
\begin{tabular}{ccccc}   
\hline\hline                                                                                                                    
Parameter & \textit{Planck}+\textit{R19} & \textit{Planck}+\textit{lensing}+\textit{R19} & \textit{Planck}+\textit{BAO}+\textit{R19} \\ \hline

$\Omega_{\rm b}h^2$ & $0.0224 \pm 0.0002$ & $0.0224 \pm 0.0002$ & $0.0223 \pm 0.0002$ \\
 
$\Omega_{\rm c}h^2$ & $<0.062$ & $<0.068$ & $0.062^{+0.020}_{-0.017}$ \\

$n_s$ & $0.966 \pm 0.004$ & $0.967 \pm 0.004$ & $0.964 \pm 0.004$ \\

$\tau$ & $0.053 \pm 0.008$ & $0.052 \pm 0.007$ & $0.053 \pm 0.008$ \\

$\xi$ & $-0.66^{+0.09}_{-0.13}$ & $-0.62^{+0.09}_{-0.14}$ & $-0.47 \pm 0.14$ \\

$H_0\,[{\rm km}\,{\rm s}^{-1}\,{\rm Mpc}^{-1}]$ & $74.0^{+1.2}_{-1.0}$ & $74.0^{+1.4}_{-1.1}$ & $71.7 \pm 1.1$ \\

\hline\hline                                                  
\end{tabular}                                                   
\caption{Constraints on selected cosmological parameters within the coupled vacuum $\xi\Lambda$CDM model, considering three dataset combinations all involving the \textit{R19} Gaussian prior on $H_0$ based on the latest local distance measurement from \textit{HST}, in combination with: the \textit{Planck} 2018 legacy dataset (without CMB lensing); the \textit{Planck} 2018 legacy dataset including the CMB lensing power spectrum reconstructed from the CMB temperature four-point function; and the \textit{Planck} 2018 legacy dataset (without CMB lensing) in addition to a combination of Baryon Acoustic Oscillation distance measurements. Constraints are reported as 68\%~C.L. intervals. In the case of $\Omega_{\rm c}h^2$ for the \textit{Planck}+\textit{R19} and \textit{Planck}+\textit{lensing}+\textit{R19} dataset combinations the quantity quoted is the 95\%~C.L. upper limit.}
\label{tab7pr19} 
\end{table}

\begin{figure*}[!h]
\centering
\begin{center}
\includegraphics[width=.45\textwidth,keepaspectratio]{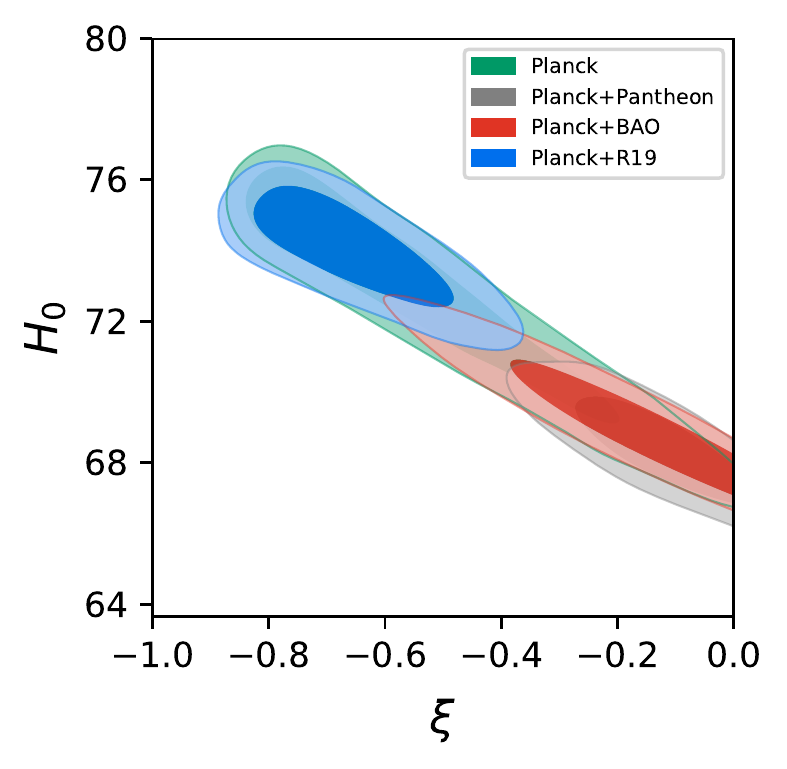} \includegraphics[width=.45\textwidth,keepaspectratio]{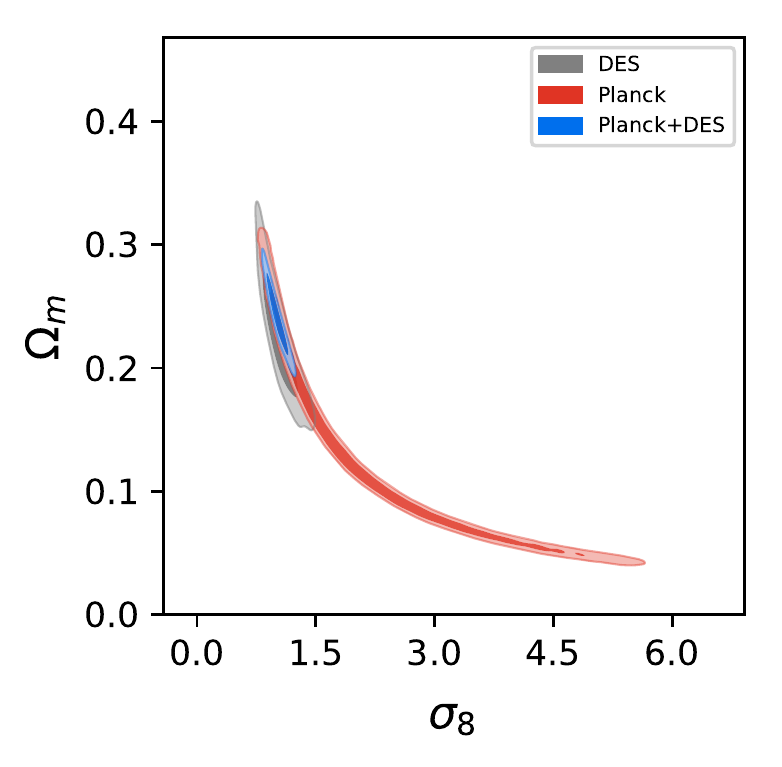}
\end{center}
\caption{Left panel: 68\% and 95\%~C.L. contours in the ($\xi$, $H_0$) plane for the \textit{Planck} (green contours), \textit{Planck}+\textit{R19} (blue contours), \textit{Planck}+\textit{BAO} (red contours), and \textit{Planck}+\textit{Pantheon} (grey contours) dataset combinations. Right panel: 68\% and 95\%~C.L. contours in the ($\sigma_8$, $\Omega_{\rm m}$) plane for the \textit{Planck} (red contours), \textit{DES} (grey contours), and \textit{Planck}+\textit{DES} (blue contours) dataset combinations.}
\label{fig}
\end{figure*}

\begin{figure}[!h]
\centering
\begin{center}
\includegraphics[height=6.5cm,keepaspectratio]{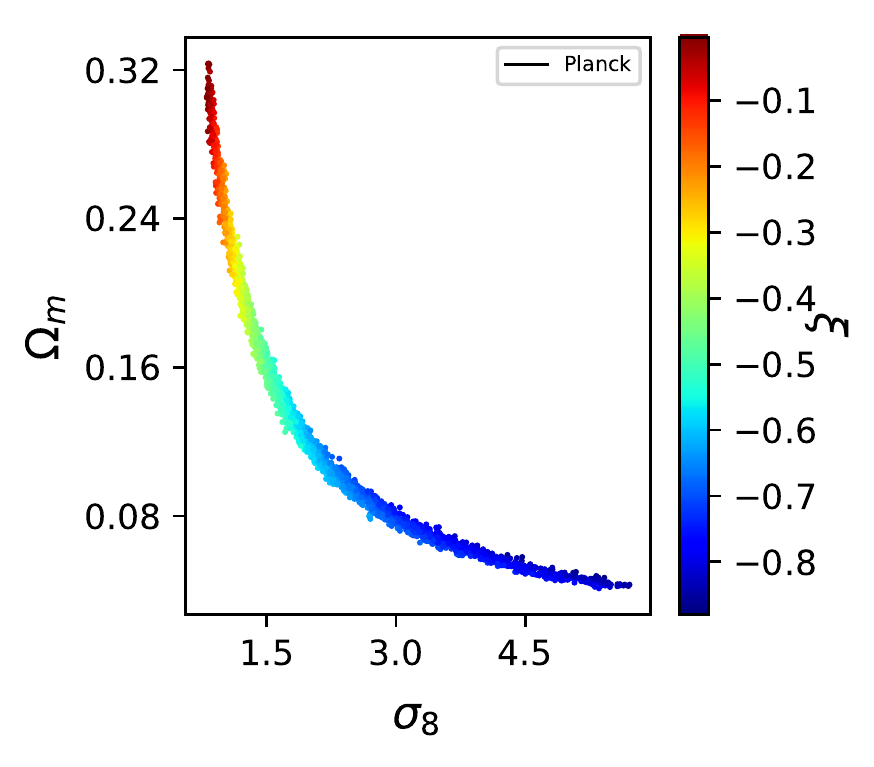} 
\end{center}
\caption{Samples in the ($\sigma_8$, $\Omega_{\rm m}$)  plane, color-coded by $\xi$, obtained from \textit{Planck} CMB data.}
\label{figxi}
\end{figure}

\section{Results}
\label{sec:results}

Our main results are shown in Tab.~\ref{tab7p}, Tab.~\ref{tab7pr19}, and Fig.~\ref{fig}. In particular, in Tab.~\ref{tab7p} we show results obtained either with the \textit{Planck} dataset alone, or combining the latter with the \textit{lensing}, \textit{BAO}, and \textit{Pantheon} datasets, one at a time. In Tab.~\ref{tab7pr19} we instead show results involving the \textit{R19} prior, considering in particular the \textit{Planck}+\textit{R19}, \textit{Planck}+\textit{lensing}+\textit{R19}, and \textit{Planck}+\textit{BAO}+\textit{R19} dataset combinations.

As shown in Tab.~\ref{tab7p}, from the \textit{Planck} dataset alone, the value of the Hubble constant $H_0$ inferred within the $\xi\Lambda$CDM model is $H_0=72.8 ^{+3.0}_{-1.5}\,{\rm km}\,{\rm s}^{-1}\,{\rm Mpc}^{-1}$. While the uncertainty is larger than that reported in~\cite{Aghanim:2018eyx} within the standard $\Lambda$CDM scenario, the central value has significantly shifted upwards. Indeed, this value is perfectly consistent with the \textit{HST}  measurement of~\cite{Riess:2019cxk}, showing an agreement well below the $1\sigma$ level. Therefore, within the $\xi\Lambda$CDM model, the $H_0$ tension is solved when considering \textit{Planck} data alone. In addition, we find $\xi=-0.54^{+0.12}_{-0.28}$, an apparent $>4\sigma$ detection of DM-DE interactions.

As we see from Tab.~\ref{tab7p}, including the \textit{lensing} dataset does not alter our conclusions. In particular, from the \textit{Planck}+\textit{lensing} dataset combination we still find a strong detection of DM-DE interactions, with $\xi=-0.51^{+0.12}_{-0.29}$, and a high value of $H_0=72.8^{+3.3}_{-1.6}\,{\rm km}\,{\rm s}^{-1}\,{\rm Mpc}^{-1}$. Including the \textit{lensing} dataset has therefore mostly enlarged the uncertainties and very slightly driven the results towards $\Lambda$CDM, but to an extent which is unable to alter our conclusions neither qualitatively nor quantitatively.

The reason for such a high value of $H_0$ from CMB measurements alone can be found in the strong degeneracy between $H_0$ and $\xi$, as depicted in the left panel of Fig.~\ref{fig}. The origin of this degeneracy resides in the fact that for the $\xi\Lambda$CDM model, the background evolution of the DM energy density  has an extra contribution proportional to the absolute value of $\xi$ and growing with $(1+z)^3$. Due to the presence of this extra term, the amount of DM today, $\Omega_{\rm c}$, must be smaller. However, the acoustic peak structure of the CMB (and in particular the relative height of odd and even peaks, as well as the overall height of all peaks) accurately fixes the value of $\Omega_{\rm c} h^2$: in order to accommodate a lower value of $\Omega_c$, a higher value of $H_0$ is required. An inverse correlation between $\xi$ and $H_0$ is therefore expected, and is reflected in the contours in the left panel of Fig.~\ref{fig}.

Note that even if the \textit{Planck} dataset alone shows a preference for a non-zero negative $\xi$ at $>95\%$~C.L.,  this is likely due to a volume effect, \textit{i.e.} more models with $\xi<0$ are compatible with \textit{Planck} than models with $\xi=0$. This explanation is supported by the fact that the best-fit $\chi^2$ for $\xi \neq 0$ is almost the same as the best-fit $\chi^2$ for $\Lambda$CDM. Computing the Bayes factor for the $\xi\Lambda$CDM model with respect to $\Lambda$CDM for the \textit{Planck} dataset we find $\ln B=1.3$. According to the modified Jeffreys scale of~\cite{Kass:1995loi}, this indicates a positive preference for the $\xi\Lambda$CDM model. When we also include the \textit{lensing} dataset, we find that the Bayes factor is reduced to $\ln B=0.9$, which still indicates a preference for the $\xi\Lambda$CDM model, but at a weak level.

As the \textit{Planck} and \textit{R19} datasets are now consistent, it is possible to combine them. When considering the \textit{Planck}+\textit{R19} combination, we find an even stronger indication for non-zero $\xi$, inferring $\xi = -0.66^{+0.09}_{-0.13}$, an apparent $>5\sigma$ detection of DM-DE interactions. The Hubble constant is instead constrained to be $H_0=74.0^{+1.2}_{-1.0}\,{\rm km}\,{\rm s}^{-1}\,{\rm Mpc}^{-1}$. Computing the Bayes factor, we find the extremely high value $\ln B=10.0$, indicating a very strong preference for the $\xi\Lambda$CDM model. Similarly, we can also consider the \textit{Planck}+\textit{lensing}+\textit{R19} dataset combination, given the mutual compatibility between \textit{Planck}+\textit{lensing} and \textit{R19}. Doing so, we find, as already seen earlier, that the \textit{lensing} dataset has mostly enlarged the uncertainties and very slightly drives the results towards $\Lambda$CDM. This is however insufficient to qualitatively or quantitatively alter our earlier conclusions, as we find $\xi=-0.62^{+0.09}_{-0.14}$ and $H_0=74.0^{+1.4}_{-1.1}\,{\rm km}\,{\rm s}^{-1}\,{\rm Mpc}^{-1}$.

As we discussed earlier, it is important to also consider BAO and SNeIa distance measurements when assessing the possibility of addressing the $H_0$ tension through late-time deviations from $\Lambda$CDM~\cite{Bernal:2016gxb,Feeney:2018mkj,Lemos:2018smw,Aylor:2018drw,Knox:2019rjx}. We therefore further investigate the impact of including late-time measurements of the expansion history from the \textit{BAO} and \textit{Pantheon} datasets. This results in all our previous findings being mildened. In fact, for the both the \textit{Planck}+\textit{BAO} and \textit{Planck}+\textit{Pantheon} dataset combinations, we infer a slightly lower value of $H_0$ within the $\xi\Lambda$CDM model than the value previously inferred from \textit{Planck} alone. In particular, from the \textit{Planck}+\textit{BAO} dataset combination we find $H_0=69.4^{+0.9}_{-1.5}\,{\rm km}\,{\rm s}^{-1}\,{\rm Mpc}^{-1}$, while from the \textit{Planck}+\textit{Pantheon} dataset combination we find $H_0=68.6^{+0.8}_{-1.0}\,{\rm km}\,{\rm s}^{-1}\,{\rm Mpc}^{-1}$. While both figures are clearly not high enough to indicate a strong resolution to the $H_0$ tension, they do nonetheless bring the tension down to the $\approx 3\sigma$ level. For the \textit{Planck}+\textit{BAO} dataset combination, in particular, the tension is brought down to the $2.6\sigma$ level, at which point one might start to view the $H_0$ tension as simply a statistical fluctuation within the $\xi\Lambda$CDM model.

For both the \textit{Planck}+\textit{BAO} and \textit{Planck}+\textit{Pantheon} dataset combinations, the hint for a non-zero coupling is still present, albeit at a much lower statistical significance, just above $1\sigma$. This is further confirmed when we compute the Bayes factor for the $\xi\Lambda$CDM model with respect to the $\Lambda$CDM model. We find $\ln B=-0.7$ for the \textit{Planck}+\textit{BAO} dataset combination, corresponding to a weak preference for $\Lambda$CDM, and $\ln B=-1.5$ for the \textit{Planck}+\textit{Pantheon} dataset combination, corresponding to a positive preference for $\Lambda$CDM. These results indicate that the increased model complexity of the $\xi\Lambda$CDM model is not warranted by a sufficient overall improvement in fit when including late-time measurements of the expansion history, unlike when the \textit{Planck} dataset alone or the \textit{Planck}+\textit{R19} dataset combination are considered.

As we see from the left panel of Fig.~\ref{fig}, the \textit{Planck}+\textit{R19} and \textit{Planck}+\textit{BAO} dataset combinations are mildly consistent, since their $2\sigma$ credible regions overlap to a non-negligible extent. Furthermore, as we saw earlier, the value of $H_0$ obtained for the \textit{Planck}+\textit{BAO} dataset combination only shows a $2.6\sigma$ tension with \textit{R19}. This means that we can consider the \textit{Planck}+\textit{BAO}+\textit{R19} dataset combination, although with caution. From this combination, we find $\xi=-0.47 \pm 0.14$ and $H_0=71.7 \pm 1.1\,{\rm km}\,{\rm s}^{-1}\,{\rm Mpc}^{-1}$. These figures are certainly intriguing and further highlight that, when the \textit{BAO} dataset is included, the $\xi\Lambda$CDM model remains an interesting contender to address the $H_0$ tension (this is no longer true if the \textit{Pantheon} dataset is included).

Overall, we find that when late-time measurements of the expansion history are included, our findings are mildened. While the $\xi\Lambda$CDM model clearly can no longer be considered an extremely promising solution to the $H_0$ tension, it is still able to bring the tension down to an interesting level, where it could be attributable (or at least almost attributable) to a statistical fluctuation. However, our findings also reinforce previous results indicating that finding late-time solutions to the $H_0$ tension is challenging (but not impossible)~\cite{Bernal:2016gxb,Feeney:2018mkj,Lemos:2018smw,Aylor:2018drw,Knox:2019rjx}. We also remind the reader that, within the $\xi\Lambda$CDM model, the \textit{Planck} and \textit{BAO}/\textit{Pantheon} datasets are in mild tension, and it is therefore not completely clear whether one is allowed to combine them in first place. In Sec.~\ref{subsec:baosne}, we discuss possible model-dependence issues pertaining to the use of BAO and SNeIa data when constraining IDE models and assessing their ability to address the $H_0$ tension.

The solution to the $H_0$ tension due to a lower intrinsic value for $\Omega_{\rm c}$ at present within the $\xi\Lambda$CDM model implies a much larger degeneracy in the $\Omega_{\rm m}-\sigma_8$ plane, reflected in the right panel of Fig.~\ref{fig}: the allowed contours from the \textit{Planck} dataset follow a band, rather than reproducing the small region usually singled out. The reason is that once a coupling $\xi$ is switched on, the required DM energy density $\Omega_{\rm c}$ must be smaller as we have seen, implying that the clustering parameter $\sigma_8$ must be larger to have a proper normalization of the (lensing and clustering) power spectra. This effect can be understood from the scatter plot in the $\Omega_{\rm m}-\sigma_8$ plane in Fig.~\ref{figxi}: as the absolute value of $\xi$ is increased, the allowed region bends towards larger (smaller) values of $\sigma_8$ ($\Omega_{\rm m}$).

The \textit{DES} contours follow the expected  $S_8 \equiv \sigma_8 (\Omega_{\rm m}/0.3)^{0.5} \simeq 0.79$  behavior~\cite{Aghanim:2018eyx} . Notice that the \textit{DES}  and \textit{Planck} contours overlap for a very large fraction of the parameter space in the $\Omega_{\rm m}-\sigma_8$ plane, implying that the tension between \textit{Planck} and \textit{DES} is alleviated. Notice that this is not merely an effect due to the larger uncertainties in the \textit{Planck} contours, but rather is due to the strong overlap between the two contours. We have found that removing one at a time the shear-shear, galaxy-galaxy, or galaxy-shear correlation functions from \textit{DES} data does not qualitatively impact our results, with the overlap between the \textit{Planck} and \textit{DES} contours in the $\Omega_{\rm m}-\sigma_8$ plane remaining substantial.

\subsection{Note on BAO and SNeIa measurements}
\label{subsec:baosne}

A comment on BAO and SNeIa measurements is in order at this point. It is worth noting that the BAO measurements have been extracted under the assumption of $\Lambda$CDM. In fact, assuming a fiducial cosmology is needed both for converting the galaxy catalogue from celestial to comoving cartesian coordinates, as well as for performing the reconstruction procedure which sharpens the BAO peak. Therefore, fiducial cosmology assumptions and possible model-dependence certainly enter at multiple stages of BAO analyses. Whether or not this makes a significant difference in models beyond $\Lambda$CDM, is a non-trivial question which is still open. Aspects of this problem have been addressed in recent works, see e.g.~\cite{Sherwin:2018wbu,Carter:2019ulk,Heinesen:2019phg,Bernal:2020vbb}.

In particular~\cite{Heinesen:2019phg} has shown that, in the presence of models which deviate significantly from $\Lambda$CDM at late times (in particular if large metric gradients are present), the usual Alcock-Paczynski scaling underlying the standard BAO measurements breaks down. In other words, the two key assumptions in BAO analyses that 1) differences in the distance-redshift relationship between true and fiducial cosmology scale linearly, and 2) differences in comoving clustering between true and fiducial cosmology can be nulled by the same free parameters used to null the non-BAO signal in the 2-point correlation function or galaxy power spectrum (see e.g.~\cite{Carter:2019ulk}), break down. As a result, it was suggested in~\cite{Heinesen:2019phg} that current BAO measurements should not be extrapolated to constrain cosmologies where late-time distance measures differ more than a few percent from $\Lambda$CDM. At face value, this issue would appear to apply to the $\xi\Lambda$CDM model, where late-time distance measures deviate at percent-level from $\Lambda$CDM if the $H_0$ tension is to be solved. Work to quantify the extent to which this possible residual BAO model-dependence affects our results, and in particular how much BAO measurements can be trusted to constrain the $\xi\Lambda$CDM model, is underway, and we hope to report on this in the near future. In the meantime, we advise caution with regards to the interpretation of results stemming from dataset combinations involving the \textit{BAO} dataset in this work.

With regards to SNeIa distance measurements, no specific model assumption was made in obtaining these. However, the interpretation thereof, or in other words the \textit{Pantheon} likelihood, assumes that intrisic SNeIa luminosities do not evolve with redshift. Whether or not this is actually valid has been the subject of recent debate (see e.g.~\cite{Kang:2019azh,Rose:2020shp}). More importantly, even if astrophysics is not responsible for redshift evolution of intrinsic SNeIa luminosities, such an evolution can occur in models of DE beyond the cosmological constant, particularly if these lead to time-dependent fundamental constants or to coupling between DE and baryons (as for instance in~\cite{Vagnozzi:2019kvw,Jimenez:2020ysu,Kase:2020hst}): see~\cite{Calabrese:2013lga,Wright:2017rsu} for examples of the impact on SNeIa luminosities.

Since we have not assumed our coupling function $Q$ to arise from any particular Lagrangian, it is hard (if not impossible) to assess what impact our IDE model should have on intrinsic SNeIa luminosities. In~\cite{DiValentino:2020evt}, a model-agnostic parametrization of a possible redshift-dependence in intrinsic SNeIa luminosities in the \textit{Pantheon} likelihood showed that for the $\xi\Lambda$CDM model in question, the effect of this possible systematic is to slightly raise $H_0$ and enlarge the uncertainties thereof (see Tabs.~6 and~7 of~\cite{DiValentino:2020evt}). This small but noticeable effect could help make the case for IDE models providing an interesting route towards addressing the $H_0$ tension even when SNeIa luminosity distance measurements are considered.

In summary, we certainly believe that the inclusion of BAO and SNeIa data when constraining late-time deviations from $\Lambda$CDM is extremely important, particularly in terms of assessing the feasibility of addressing the $H_0$ tension within these models. However, care must be taken with regards to the interpretation of these datasets and their use in constraining models beyond $\Lambda$CDM. This is the case for BAO measurements if late-time distance measures differ more than a few percent from $\Lambda$CDM in the target cosmology, or for SNeIa measurements if the DE component in the target cosmology couples to baryons and/or leads to time-dependent fundamental constants.

\section{Conclusions}
\label{sec:conclusions}

In this work, we have examined the persisting $H_0$ tension in light of the \textit{Planck} 2018 legacy data release, late-time distance measurements of the expansion history from BAO and SNeIa, and the latest $1\%$ determination of $H_0$ from \textit{HST}. We find that within an interacting dark energy model studied in previous works, with coupling function given by Eq.~(\ref{eq:coupling}), the value of $H_0$ inferred by \textit{Planck} is consistent with the latest local distance measurement well within $1\sigma$, representing a solution to the $H_0$ tension.

Our findings are, however, mildened when we include late-time BAO and SNeIa measurements. In this case, the value of $H_0$ we infer is not high enough that we can claim a definitive solution to the $H_0$ tension. Nonetheless, the tension is brought down to a level where it could be compatible with a statistical fluctuation within the interacting dark energy model we have considered, although the model is not favoured by Bayesian evidence considerations. At the same time, our interacting dark energy model appears promising in terms of alleviating the tensions between CMB and cosmic shear measurements. In particular, we observe a considerably improved overlap between the \textit{Planck} and \textit{DES} contours in the $\Omega_{\rm m}-\sigma_8$ plane.

Possible model-dependence issues pertaining to the use of BAO and SNeIa data when constraining interacting dark energy models and assessing their ability to address the $H_0$ tension are discussed in Sec.~\ref{subsec:baosne}. It is certainly very important to include BAO and SNeIa measurements, from which an inverse distance ladder can be constructed, when constraining late-time deviations from $\Lambda$CDM. Nonetheless, we feel like advising slightly more caution when interpreting our results involving either or both the \textit{BAO} or \textit{Pantheon} datasets, due to these possible model-dependence issues we have discussed, and to their tension with the \textit{Planck} dataset when interpreted within the framework of an interacting dark energy model.

It is worth keeping in mind that we have not varied the dark energy equation of state but fixed it to $w=-0.999$ (which in turn requires $\xi<0$), an approach already followed in the earlier~\cite{Salvatelli:2013wra,DiValentino:2017iww}. As explained at the start of Sec.~\ref{sec:methodology}, we did so to provide a surrogate for a coupled vacuum model, while bypassing the usual instabilities problem of the latter~\cite{Valiviita:2008iv}. The case where $w$ is varied as well was considered in our later companion paper~\cite{DiValentino:2019jae}, and shown to decrease the Bayes factor in favour of the resulting IDE model. It is also worth recalling that the dark matter-dark energy coupling we have studied is purely phenomenological, introduced at the level of the continuity equations, and not coming from a specific Lagrangian.

To conclude, it is intriguing that the interacting dark energy model we have considered provides not only an interesting solution to the $H_0$ tension, but at the same time can alleviate the tension between CMB and cosmic shear measurements. Our findings suggest that the possibility of late-time new physics in the dark sector hiding behind these persisting cosmological tensions cannot yet be discarded. Future data (e.g. from \textit{Euclid} or from gravitational wave standard sirens) will be vital towards further confirming the trends we have seen, or definitively discarding the model~\cite{Yang:2019vni,Lucca:2020zjb}.

\section*{Acknowledgements}
E.D.V. acknowledges support from the European Research Council in the form of a Consolidator Grant with number 681431. A.M. is supported by TASP, iniziativa specifica INFN. O.M. is supported by PROMETEO II/2014/050, by the Spanish Grant FPA2017-85985-P of the MINECO, by the MINECO Grant SEV-2014-0398 and by the European Union’s Horizon 2020 research and innovation programme under the Marie Sk\l odowska-Curie grant agreements 690575 and 674896. S.V. is supported by the Isaac Newton Trust and the Kavli Foundation through a Newton-Kavli fellowship, and acknowledges a College Research Associateship at Homerton College, University of Cambridge. O.M. would like to thank the hospitality of the Fermilab Theory Department. This work is based on observations obtained with Planck (www.esa.int/Planck), an ESA science mission with instruments and contributions directly funded by ESA Member States, NASA, and Canada. We acknowledge the use of the Planck Legacy Archive. 

\bibliographystyle{JHEP}
\bibliography{IDE.bib}

\end{document}